\documentclass[pre,twoside,twocolumn,showpacs,byrevtex,superscriptaddress]{revtex4}

\newcommand{\Arect}{A_{\mbox{\scriptsize rect}}}
\newcommand{\Srect}{L_{\mbox{\scriptsize rect}}}

\newcommand{\hatalpharect}{\hat{\alpha}_{\mbox{\scriptsize rect}}}

\newcommand{\Aspoke}{A_{\mbox{\scriptsize $n$-spoke}}}
\newcommand{\Sspoke}{L_{\mbox{\scriptsize $n$-spoke}}}

\newcommand{\hatalphanspoke}{\hat{\alpha}_{\mbox{\scriptsize $n$-spoke}}}

\newcommand{\Aeight}{A_{\mbox{\scriptsize 8-point}}}
\newcommand{\Seight}{L_{\mbox{\scriptsize 8-point}}}

\newcommand{\hatalphaeightstar}{\hat{\alpha}_{\mbox{\scriptsize 8-point}}}

\newcommand{\Prn}{P(r; n)}

\newcommand{\Pins}{P_{\rm ins}(r; n)}

\usepackage{graphicx,epsfig,verbatim,enumerate}
\usepackage{amssymb}
\usepackage{ifthen}
\newboolean{twocolswitch}

\newcommand{\www}[1]{\url{#1}}
\newcommand{\req}[1]{(\ref{#1})}

\setboolean{twocolswitch}{true}

\begin{document}

\title{
Packing-limited growth of irregular objects
}

\markboth{PACKING LIMITED GROWTH}{PETER SHERIDAN DODDS AND JOSHUA S.\ WEITZ}

\author{
  \firstname{Peter Sheridan}
  \surname{Dodds}
  }
\email{peter.dodds@columbia.edu}
\affiliation{
        Institute for Social and Economic Research and Policy,
        Columbia University,
        New York, NY 10027.
        }

\author{
  \firstname{Joshua S.}
  \surname{Weitz}
  }
\email{jsweitz@segovia.mit.edu}
\thanks{(Please direct correspondence to both authors.)}
\affiliation{
        Department of Earth, 
        Atmospheric and Planetary Sciences,
        Massachusetts Institute of Technology, 
        Cambridge, MA 02139.
        }
\affiliation{
        Department of Physics,
        Massachusetts Institute of Technology,
        Cambridge, MA 02139.
        }

\date{\today}

\begin{abstract}
We study growth limited by packing for irregular
objects in two dimensions.   We generate packings
by seeding objects randomly in time and space
and allowing each object to grow until it collides
with another object.  
The objects we consider
allow us to investigate the separate effects of
anisotropy and non-unit aspect ratio.
By means of a connection to the decay of pore-space volume,
we measure power law exponents for the object size distribution.
We carry out a scaling analysis,
showing that it provides an upper bound for the size distribution exponent.
We find that while the details of the growth mechanism are irrelevant, 
the exponent is strongly shape dependent.
Potential applications lie in ecological and biological
environments where sessile organisms compete for
limited space as they grow.
\end{abstract}


\maketitle

\section{Introduction}
\label{irregular.introduction}

In a previous work~\cite{dodds2002pa},
we examined packings formed by
spheres growing in $d$ dimensions,
immediately stopping upon contact with another sphere.
We termed this to be packing-limited growth (PLG).
Here, we address the question of what happens
when we consider irregular (i.e., non-spherical) objects 
of similar shape growing in $d=2$ dimensions. 
We focus in particular on collisions of growing rectangles.  
By combining rectangles, we
are able to form and examine a range of shapes whose growth
patterns vary broadly in radial anisotropy.
Whereas in~\cite{dodds2002pa} we found universality classes
depending only on dimension $d$, we find here
that marked non-universal behavior arises when shapes are varied.

Our problem finds several motivations.
First, this is a relatively unexplored kind of packing.
Packings are generally static or randomly generated~\cite{mandelbrot1977}: 
physical mechanisms are only seldomly connected with
the creation of packings~\cite{brilliantov94,andrienko94}.
Furthermore, packings are typically of monodisperse objects
of the same form or taken from a small set of forms~\citep{huillet1998,conway1999}.
From a physical point of view,
two-dimensional packings of growing
objects may be of use in modeling or understanding
certain biological and ecological patterns.  
We consider the geometric approach presented
here as a preliminary step towards describing how shape 
alters the size distributions of populations.

In section~\ref{irregular.objects}
we describe the various objects we
construct from rectangles.  Although
not all of these objects have obvious
physical parallels, they present a range
of typological cases from which direct
applications and comparisons may be sought.
In section~\ref{irregular.theory}, we 
provide a scaling analysis that
empirically appears to be exact for 
$d \ge 4$ in the
case of hyperspheres~\cite{dodds2002pa}
(details of the calculations for this
section are given
in Appendix~\ref{appendix:irregular.alphamf}).
We report the results of our numerical
investigations in section~\ref{irregular.numerics}
along with a discussion of the failings of
the scaling theory.
We conclude the paper in section~\ref{irregular.conclusion}
and outline our algorithm for packing rectangles 
in Appendix~\ref{appendix:irregular.rectcalcs}.

\section{Irregular objects considered}
\label{irregular.objects}

\begin{figure}[tbhp!]
  \begin{center}
    \epsfig{file=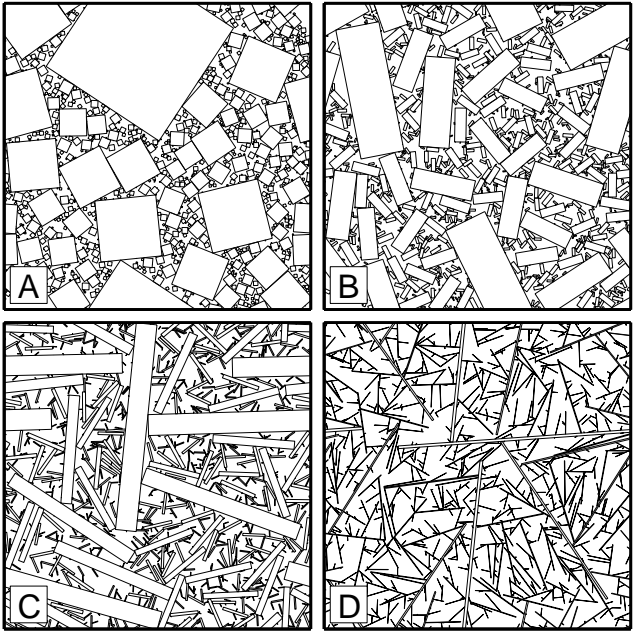,width=1\columnwidth}
    \caption{
      Rectangle packings 
      of the unit square for varying aspect ratio $a$
      created using the packing-limited growth mechanism.
      Each packing consists of 1000 rectangles 
      with periodic boundary conditions being imposed.
      The aspect ratios corresponding to the packings
      are (A) $a=1$, (B) $a=3$, (C) $a=10$, and 
      (D) $a=100$.  The first three
      packings are initialized with
      four randomly placed and oriented
      rectangles with longest side length 0.25,
      with eight such rectangles being used in the fourth
      packing.
      }
    \label{fig:irregular.rectsN1000comb}
  \end{center}
\end{figure}

\begin{figure}[tbhp!]
  \begin{center}
    \epsfig{file=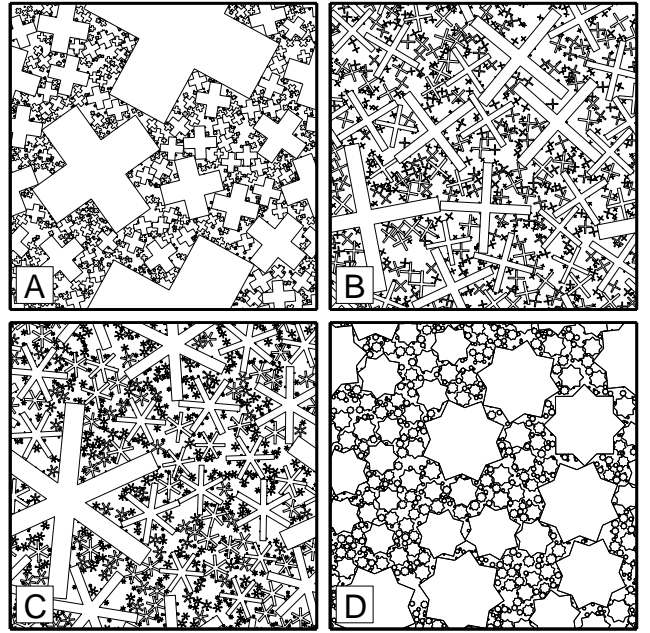,width=1\columnwidth}
    \caption{
      Packings of irregular objects 
      formed by combinations of rectangles.
      As per Figure~\ref{fig:irregular.rectsN1000comb},
      each packing contains 1000 objects
      and the boundary conditions are periodic.
      In (A) and (B),
      the crosses are formed by two rectangles at
      right angles with aspect ratios 
      $a=3$ and $a=10$ respectively.
      Plot (C) shows a packing of six-legged
      stars composed of three rectangles with
      aspect ratio $a=10$.
      In (D), the packing object is an
      eight-pointed star formed by 
      two overlapping squares set at an angle of
      $\pi/4$ to each other.
      }
    \label{fig:irregular.crossesN1000comb}
  \end{center}
\end{figure}

The building block shape of the objects we consider is the
rectangle.  We denote the
aspect ratio as $a$, defining all rectangles such that $a \ge 1$.
We record rectangle size $r$ as half the length of the long side,
the dimensions then being $2r$ and $2r/a$.
Examples of each object are provided in the packings
of Figures~\ref{fig:irregular.rectsN1000comb}
and~\ref{fig:irregular.crossesN1000comb}.

For $a=1$, we have squares and as $a\rightarrow \infty$,
rectangles effectively become line segments and the packing
becomes one of fitting $d=1$ objects into a $d=2$ volume.
Rectangles afford a basic example of anisotropic growth
since the long side grows at a rate slower than the short
side by a factor of $1/a$.  
The growing square is distinguished
from the rectangle since the former
expands uniformly perpendicular to its edges.
However, in comparison to disks,
the growth rate of edge points relative to the center of
both squares and rectangles is non-uniform.

The simplest combination of two rectangles is a cross 
(Figures~\ref{fig:irregular.crossesN1000comb}A
and~\ref{fig:irregular.crossesN1000comb}B)
and a natural generalization is the $2n$-spoke
object (with $n=1$ being a rectangle and $n=2$ being a cross, 
see Figure~\ref{fig:irregular.crossesN1000comb}C).
Each $2n$-spoke object is then part of a family of shapes indexed by $a$.

The last object we consider is an eight-pointed star
as shown in Figure~\ref{fig:irregular.crossesN1000comb}D.
This object is formed by two squares overlaid at angle
of $\pi/4$ to each other.  

All objects are packed using the approach of
Manna~\cite{manna92a}: objects are added sequentially
and allowed to instantly grow so as to just reach
the existing packing structure.
We have observed and argued~\cite{dodds2002pa} that
the value of the size distribution exponent and all other
related exponents is independent of the growth dynamic.

In the context of packing,
spheres make for straightforward calculations since
the contact point between any two colliding spheres 
occurs along the line through their centers.
Rectangles constitute a relatively simple generalization
of spheres from a numerical point of view,
hence our use of them here.

\section{Scaling theory}
\label{irregular.theory}

\begin{figure}[tbhp!]
  \begin{center}
    \epsfig{file=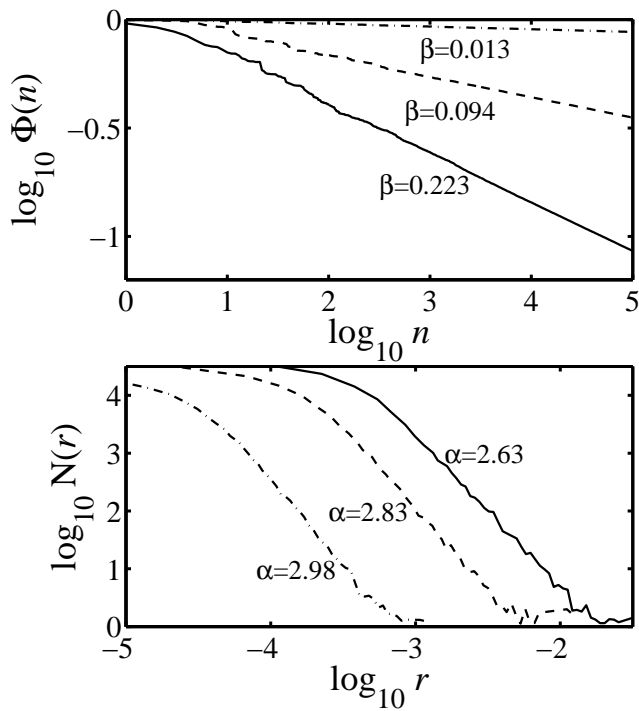,width=1\columnwidth}
    \caption{
      For rectangle packings of $10^5$ objects,
      plots of $\Phi(n)$, the decay of pore space as a function
      of number of rectangles added $n$, and $N(r)$, the frequency of
      objects of size $r$, for rectangles.  In both plots,
      the rectangles have aspect ratios $a=1$ (solid line),
      $a=10$ (dashed line) and $a=100$ (dot-dash line).
      $N(r)$ is binned in log-space for clarity.
      The exponent $\alpha$ (in $N(r) \propto r^{-\alpha}$)
      increases with aspect ratio $a$, limiting to $3$ as
      $a \rightarrow \infty$.  Correspondingly,
      $\beta$ decreases towards 0 (see equation~\req{eq:irregular.alphabeta}).
      }
    \label{fig:irregular.rectsphi}
  \end{center}
\end{figure}

In~\cite{dodds2002pa}, we argued that a simple scaling
assumption may be made regarding the form of $\Prn$,
the distribution of sphere sizes after $n$ objects
have been packed.  We take $\Prn$ to be described
by a power law for radii above a cutoff value $r_c$
and uniform below:
\begin{equation}
  \label{eq:irregular.Prn}
  \Prn = 
  \left\{
    \begin{array}{ll}
      \frac{\alpha-1}{\alpha} r_c^{-1} & \mbox{for $r<r_c$}\\
      \frac{\alpha-1}{\alpha} r_c^{-1} 
      \left(\frac{r}{r_c} \right)^{-\alpha} & \mbox{for $r \geq r_c$}.
    \end{array}
  \right.
\end{equation}
The tail of the distribution is fixed 
and the distribution fills in ($r_c$ decreases with $n$)
with the $(n+1)$th object being chosen from within the
pore space.  The size distribution $\Prn$
is connected to the probability of inserting an object
of size $r$ after $n$ objects have been deposited, $\Pins$.
Since the probability
of adding a sphere of vanishing radii must be proportional
to the total surface area of the existing spheres $S(n)$,
we were able to estimate $\Pins$ as~\cite{dodds2002pa}
\begin{equation}
  \label{eq:irregular.Pins}
  \Pins  = \frac{S(n)}{\Phi(n)}, \quad 0 \le r \le r_c,
\end{equation}
where $\Phi(n)$ is the pore space volume.
Using equation~\req{eq:irregular.Prn} to
calculate $S(n)$ and $\Phi(n)$ and requiring
that $\Pins$ be normalized, we obtained the estimate
$\alpha=11/4$ for disks.

More generally in $d=2$ dimensions, if we now 
write area as $A=\rho r^2$ and perimeter as $L=\sigma r$,
we find the scaling exponent to be
\begin{equation}
  \label{eq:irregular.alphamf}
  \hat{\alpha} = \frac{9\sigma + 4\rho}{3\sigma + 2\rho}.
\end{equation}
Setting $\sigma=2\pi$ and $\rho=\pi$,
we recover $\alpha=11/4$ for disks.
Note that the probability of adding a non-spherical
object is still proportional to $S(n)$ in the limit of
$r \rightarrow 0$, regardless of the object's shape.
For example,
it is irrelevant 
that an added square of side length $2r$ may be
oriented such that its center is between $r$ and $\sqrt{2}r$
away from its point of contact with the existing packing.
Also note that as $a\rightarrow\infty$, the theoretical
prediction tends to $\hat{\alpha}=3$, in agreement with
a previous determination of an upper bound for
polydisperse packings~\cite{aste1996}.  
Details of the calculations finding $\hat{\alpha}$
for the objects we consider here are to be found in
Appendix \ref{appendix:irregular.alphamf}.

All other exponents depend on $\alpha$ via
simple scaling relations~\cite{dodds2002pa}.
In particular, the pore volume $\Phi(n)$ 
decays as $n^{-\beta}$ with the connection
between $\beta$ and $\alpha$ being
\begin{equation}
  \label{eq:irregular.alphabeta}
  \alpha = 1 + \frac{2}{1+\beta}.
\end{equation}
We use this equation to
calculate theoretical estimates of $\beta$,
denoting them by $\hat{\beta}$.
Since direct measurement of $\beta$ is a significantly
more robust exercise than determining $\alpha$ from $P(r)$,
we employ equation~\req{eq:irregular.alphabeta}
in estimating $\alpha$ in the following section.

\section{Numerical results}
\label{irregular.numerics}

\begin{table}[tbp!]
 \begin{center}
   \begin{tabular*}{\columnwidth}{@{\ }l@{\extracolsep{\fill}}ccccc@{\ }}
     \toprule 
     object & $\beta$ & $\hat{\beta}$ & $\alpha$ & $\hat{\alpha}$ \\ 
     \colrule
     disk                     & 0.278(1)  & 0.1429  & 2.564(1) & 2.750   \\
     square                   & 0.223(2)  & 0.1429  & 2.635(2) & 2.750   \\
     rectangle ($a=2$)        & 0.207(2)  & 0.1000  & 2.656(2) & 2.818   \\
     rectangle ($a=5$)        & 0.145(2)  & 0.0526  & 2.746(2) & 2.900   \\
     rectangle ($a=10$)       & 0.094(1)  & 0.0294  & 2.828(2) & 2.943   \\
     rectangle ($a=20$)       & 0.055(1)  & 0.0156  & 2.897(2) & 2.969   \\
     rectangle ($a=50$)       & 0.0242(3) & 0.0065  & 2.953(1) & 2.987   \\
     rectangle ($a=100$)      & 0.0125(1) & 0.0033  & 2.975(1) & 2.993   \\
     cross ($a=3$)            & 0.169(3)  & 0.0847  & 2.710(3) & 2.844   \\
     cross ($a=10$)           & 0.074(1)  & 0.0307  & 2.862(2) & 2.940   \\
     cross ($a=100$)          & 0.009(1)  & 0.0033  & 2.983(2) & 2.993   \\
     six spoke ($a=10$)       & 0.078(2)  & 0.0318  & 2.855(3) & 2.938   \\
     eight-pointed star       & 0.213(2)  & 0.1429  & 2.648(4) & 2.750   \\
     \botrule
   \end{tabular*}
   \caption{
     Estimates of $\alpha$, the exponent of the number
     distribution, $P(r) \propto r^{-\alpha}$, for irregular objects
     undergoing packing-limited growth.
     The results for disks are included for comparison~\cite{dodds2002pa}.
     All other objects are combinations of rectangles 
     (see Figures~\ref{fig:irregular.rectsN1000comb}
     and~\ref{fig:irregular.crossesN1000comb})
     with $a$ being the aspect ratio.
     The exponent $\beta$ is determined from $\Phi(n)$
     and $\alpha$ is subsequently obtained 
     using equation~\req{eq:irregular.alphabeta}.
     Full details of the method of measuring $\beta$ are given in the text.
     Each measurement is for a single packing containing $10^5$ objects.
     Measurement errors reflect variation 
     in the approach to a limiting value of the exponents
     using the method of 
     Section~\ref{irregular.numerics}.
     The scaling theory estimates of $\hat{\beta}$ and $\hat{\alpha}$,
     which are lower and upper bounds respectively,
     are calculated using 
     equations~\req{eq:irregular.alphamf},
     \req{eq:irregular.alphabeta},
     and those given in Appendix~\ref{appendix:irregular.alphamf}.
     }
   \label{tab:irregular.alpha_results}
 \end{center}
\end{table}

For each shape, we generate statistics 
for single packings with $10^5$ objects.
Some example distributions taken from 
rectangle packings for $a=1$, 10, and 100 
are shown in Figure~\ref{fig:irregular.rectsphi}.
The distribution $N(r)$ is the unnormalized
frequency distribution corresponding to $P(r)$.
In the case of plain rectangles, we see
that as $a$ increases, both 
the observed $\alpha$ and the theoretical
estimate $\hat{\alpha}$ increase
while $\beta$ and $\hat{\beta}$ accordingly decrease.
Furthermore, for all the shapes considered we find that
$\alpha < \hat{\alpha}$ for all finite $a$.

\begin{figure}[tbp!]
  \begin{center}
    \epsfig{file=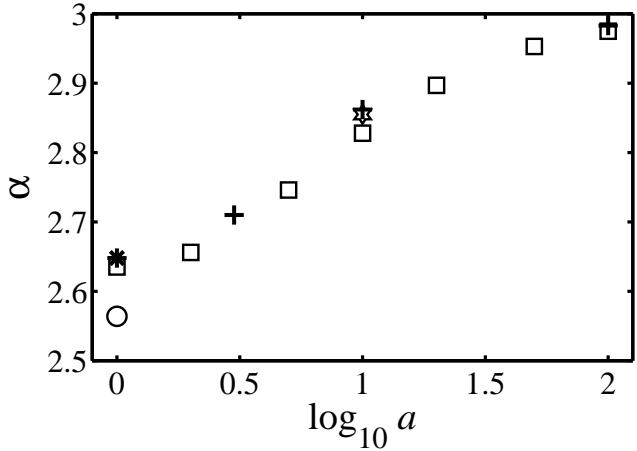,width=1\columnwidth}
    \caption{
      Measured values of $\alpha$ as a function of aspect ratio $a$ and shape.
      The symbols correspond to disks (circles), rectangles (squares), 
      crosses (plus sign), six-spoke (hexagon), and eight-pointed star
      (asterisk).
      }
    \label{fig:irregular.alphavals}
  \end{center}
\end{figure}

We note that measuring power law exponents 
is not a trivial procedure and here we take
some care to ensure the validity of our results.
While simple regression is the basic tool
of analysis, the presence of, for example, crossovers and finite
size cutoffs can substantially degrade the level of precision
attainable.  The method of measuring exponents we use here
is based on examining a smoothed version of the derivative
of the distribution as viewed in double logarithmic space
(similar approaches are to be found in~\cite{voigt97,dodds2001pa}).
For example, for $\Phi(n)$, we perform regression analysis
on $\log_{10} \Phi(n)$ versus $\log_{10}{n}$ over a sliding, variable width window
of values of $\log_{10}{n}$.  
Writing the upper and lower limits of this window 
as $\log_{10}{n_1}$ and $\log_{10}{n_2}$, we have $w = \log_{10}{n_2/n_1}$
being the width. 
In general, depending on the number of orders of magnitude spanned by the data,
we would preferably choose $w$ in the range $0.5 \le w \le 3$.
Here, we fix $w=1$ and calculate the ``local'' exponent $\beta(n_1)$ for each window.
We find in all cases that $\beta(n_1)$ tends towards a constant
value.  This indicates the scaling law is robust and
further allows us to estimate $\beta$ along with an error based
on the fluctuations observed for $\beta(n_1)$.
Measured values of $\alpha$ and $\beta$
are recorded in Table~\ref{tab:irregular.alpha_results}.

As we have noted above, our scaling theory approach
appears to be exact for $d\ge 4$ dimensions in the
case of hyperspheres and an overestimate
of the true value of $\alpha$ for $d<4$.
We therefore do not expect the scaling theory to
be correct for non-spherical objects in $d=2$ dimensions.
Indeed, in all cases, we observe 
the theoretical estimate $\hat{\alpha}$
is an overestimate of the measured $\alpha$.
This direction of error makes sense in
light of the scaling assumption made by equation~\req{eq:irregular.Pins}.
The actual form of $\Pins$ is not precisely uniform but
rather rolls over less steeply than a step function around $r=r_c$.
Given the manipulations that leads to~\req{eq:irregular.alphamf},
it can be argued that $\hat{\alpha} \ge \alpha$ must hold~\cite{dodds2002pa}.

In Figure~\ref{fig:irregular.alphavals}, we show how $\alpha$ varies
as a function of both aspect ratio $a$ of the constituent rectangles
and the object's particular shape.  
The strongest influence is evidently aspect ratio with $\alpha$ varying from
$2.564$ (disks, $a=1$) up to $3$ (rectangles, $n$-spokes, $a=\infty$).
We observe a secondary effect due to the details of the shape.
Square and the eight-pointed star packings have a value of
$\alpha$ increased above that of disks.
Cross packings have a higher value of $\alpha$ than do
rectangles with the same aspect ratio.  

\section{Concluding remarks}
We have extended a model describing the interaction
of growing disks~\cite{dodds2002pa} to the problem
of growing irregular and anisotropic objects.
In this model, the exponent characterizing
the size distribution of objects is found to be 
independent of growth dynamics. However,
the main result presented here is that the 
exponent is highly shape-dependent, adopting 
a continuous range of values $2.564\leq \alpha\leq 3$.
Ultimately, understanding how
geometry impacts on the structure of plant communities
will require imposing a notion of packing-limited growth onto a reasonable
set of dynamics.  The results here demonstrate that in so doing
we must keep in mind that shape matters.
\label{irregular.conclusion}

\begin{acknowledgments}
PSD acknowledges the support of the Columbia Earth Institute.
\end{acknowledgments}

\appendix

\appendix

\section{Shape parameters for mean field calculations}
\label{appendix:irregular.alphamf}

In this section, we derive the formulas for the area
and perimeter of irregular objects in $d=2$.  
For general rectangles, we have area given by
\begin{equation}
  \label{eq:irregular.area_rect}
  \Arect = 4r^2/a,
\end{equation}
and perimeter as
\begin{equation}
  \label{eq:irregular.perimeter_rect}
  \Srect = 4(1+1/a)r.
\end{equation}
Using equation~\req{eq:irregular.alphamf},
we therefore have the mean field estimate
of $\alpha$ for rectangles as
\begin{equation}
  \label{eq:irregular.rectmf}
  \hatalpharect = 3 - \frac{2}{5+3a},
\end{equation}
where writing the result in this fashion
makes plain the limiting value of $\hatalpharect = 3$
for $a \rightarrow \infty$.
For the $2n$-spoke objects
described above, area grows as
\begin{equation}
  \label{eq:irregular.area_spokes}
  \Aspoke = 2n/a^2(2a - \cot{\pi/2n}) r^2,
\end{equation}
and perimeter is given by
\begin{equation}
  \label{eq:irregular.perimeter_spokes}
  \Sspoke = 4n/a(a+1-\cot{\pi/2n})r,
\end{equation}
leading to the estimate 
\begin{equation}
  \label{eq:irregular.spokemf}
  \hatalphanspoke = \frac{13 + 9a -(9+2/a) \cot{\pi/2n}}{ 5 + 3a - (3+1/a)\cot{\pi/2n}}.
\end{equation}
Note that for $n=1$, we recover the
result for rectangles in equation~\req{eq:irregular.rectmf}.
Finally, for the eight-pointed star we have
\begin{equation}
  \label{eq:irregular.area_8pt}
  \Aeight =
  2\left[
    2 + (1-\tan{\pi/8})^2
  \right]r^2,
\end{equation}
and
\begin{equation}
  \label{eq:irregular.perimeter_8pt}
  \Seight = 
    16(1-\tan{\pi/8}) r,
\end{equation}
which yields
\begin{equation}
  \label{eq:irregular.8_ptmf}
  \hatalphaeightstar = 
  \frac{42 - 40 \tan{\pi/8} + 2 \tan^2{\pi/8}}
  {15 - 14 \tan{\pi/8} + \tan^2{\pi/8}} = 11/8,
\end{equation}
the same as for disks and squares.

\section{Rectangle collisions}
\label{appendix:irregular.rectcalcs}

We describe an arbitrary rectangle in
the $x$-$y$ plane as follows.  
The sides of the
rectangle are $2r$ and $2r/a$ where $r>0$
is half the length of the long side of
the rectangle and
$a \ge 1$ is take the aspect ratio.
The rectangle
is centered at $(x_0,y_0)$ and rotated
at an angle $\theta$.  We take $\theta$ as the
angle between the direction
of the positive $x$-axis and the short axis
of the rectangle (i.e., parallel to the
side with length $2r/a$) so that $0 \leq \theta < \pi$.

A rectangle described by $(r,a)$ and $\theta=0$ and $(x_0,y_0)=(0,0)$
satisfies the equation
\begin{equation}
  \label{eq:irregular.recteq}
  \max 
  \left|
    \left[
      \begin{array}{c}
        ax \\
        y
      \end{array}
    \right]
  \right|
  =
  \max 
  \left|
    \left[
      \begin{array}{cc}
        a & 0 \\
        0 & 1 \\
      \end{array}
    \right]
    \left[
      \begin{array}{c}
        x \\
        y \\
      \end{array}
    \right] 
  \right|
  = r.
\end{equation}
This is a remapping of a rectangle into a square of side length $2r$.
For an arbitrary rectangle, we can map
it onto a basic square by
recentering it at the origin, removing the rotation,
and undoing the dilation by the aspect ratio $a$:
\begin{equation}
  \label{eq:irregular.recteq2}
  \max \left|
    \left[
      \begin{array}{cc}
        a & 0 \\
        0 & 1 \\
      \end{array}
   \right]
    \left[
      \begin{array}{cc}
        \cos\theta & \sin\theta \\
        -\sin\theta & \cos\theta \\
      \end{array}
   \right]
    \left[
      \begin{array}{c}
        x-x_0 \\
        y-y_0 \\
      \end{array}
   \right]
    \right| = r.
\end{equation}
Therefore, to determine whether or not an
arbitrary point $(x,y)$ lies on or within a given
rectangle, we need to check whether or not
\begin{equation}
  \label{eq:irregular.recteq3}
  \max \left|
    \left[
      \begin{array}{cc}
        a & 0 \\
        0 & 1 \\
      \end{array}
   \right]
    \left[
      \begin{array}{cc}
        \cos\theta & \sin\theta \\
        -\sin\theta & \cos\theta \\
      \end{array}
   \right]
    \left[
      \begin{array}{c}
        x-x_0 \\
        y-y_0 \\
      \end{array}
   \right]
    \right| \leq r.
\end{equation}
In packing growing rectangles, the above
is used to check if a newly seeded rectangle
has been placed in pore space and 
not within an existing rectangle.

For a rectangle that passes this test,
the next calculation is to determine
how large it may grow preserving
its aspect ratio and orientation
so that it just reaches
an existing rectangle.

To do this, we consider one arbitrary rectangle that
does not cover the origin and find the size of
largest rectangle centered at the origin with
the short side along the $x$-axis such that
the rectangles just touch.  We will then generalize
to any configuration by appropriate rotations.

First, the ``growing'' rectangle could hit the already
existing one at any of the latter's corner points.
The four corner points are given by
\begin{eqnarray}
  \label{eq:irregular.fourcorners}
    \left[
      \begin{array}{c}
        x_0 \\
        y_0 \\
      \end{array}
   \right]
   +
   r \left[
      \begin{array}{c}
        \pm 1/a \cos\theta - \sin\theta \\
        \pm 1/a \sin\theta + \cos\theta \\
      \end{array}
   \right],
   \\
    \left[
      \begin{array}{c}
        x_0 \\
        y_0 \\
      \end{array}
   \right]
   +
   r \left[
      \begin{array}{c}
        \mp a \cos\theta + \sin\theta \\
        \mp a \sin\theta - \cos\theta \\
      \end{array}
   \right].
\end{eqnarray}

The other possible collisions
are between the corners of the added rectangle
and the sides of the existing rectangle.
There may be 0, 1, 2, 3 or 4 such interceptions
(1 or 3 if the rectangles touch at corners).
To calculate these points, we parametrize
each side of the existing rectangle and
find the interceptions with $y=\pm ax$.
For an arbitrary line segment described by
\begin{equation}
  \label{eq:irregular.xyparam}
  \begin{array}{c}
    x = b_1 + b_2 t \quad \mbox{and} \quad y = c_1 + c_2 t, \\
  \end{array}
\end{equation}
with $-1 \le t \le 1$, the intersection
with $y=\pm ax$ occurs when 
\begin{equation}
  \label{eq:irregular.xyparamint}
  \begin{array}{c}
    t = \frac{-c_1 \pm a b_1}{c_2 \mp a b_2}.
  \end{array}
\end{equation}
Parametrizing the sides of an
arbitrary rectangle gives the following:
\begin{eqnarray}
  \label{eq:irregular.sideparam1}
  \left[
    \begin{array}{c}
      x \\
      y \\
  \end{array}
  \right]
  = 
  \left[
    \begin{array}{c}
      x_0 \\
      y_0 \\
  \end{array}
  \right]
  +r
  \left[
    \begin{array}{c}
      -\sin\theta \\
      \cos\theta \\
  \end{array}
  \right]
  +\frac{rt}{a}
  \left[
    \begin{array}{c}
      \cos\theta \\
      \sin\theta \\
  \end{array}
  \right],
   \\
  \label{eq:irregular.sideparam2}
   \left[
    \begin{array}{c}
      x \\
      y \\
  \end{array}
  \right]
  = 
  \left[
    \begin{array}{c}
      x_0 \\
      y_0 \\
  \end{array}
  \right]
  +r
  \left[
    \begin{array}{c}
      \sin\theta \\
      -\cos\theta \\
  \end{array}
  \right]
  +\frac{rt}{a}
  \left[
    \begin{array}{c}
      \cos\theta \\
      \sin\theta \\
  \end{array}
  \right],
   \\
  \label{eq:irregular.sideparam3}
  \left[
    \begin{array}{c}
      x \\
      y \\
  \end{array}
  \right]
  = 
  \left[
    \begin{array}{c}
      x_0 \\
      y_0 \\
  \end{array}
  \right]
  +\frac{r}{a}
  \left[
    \begin{array}{c}
      \cos\theta \\
      \sin\theta \\
  \end{array}
  \right]
  +rt
  \left[
    \begin{array}{c}
      -\sin\theta \\
      \cos\theta \\
  \end{array}
  \right],
   \\
  \label{eq:irregular.sideparam4}
  \left[
    \begin{array}{c}
      x \\
      y \\
  \end{array}
  \right]
  = 
  \left[
    \begin{array}{c}
      x_0 \\
      y_0 \\
  \end{array}
  \right]
  +\frac{r}{a}
  \left[
    \begin{array}{c}
      -\cos\theta \\
      -\sin\theta \\
  \end{array}
  \right]
  +rt
  \left[
    \begin{array}{c}
      -\sin\theta \\
      \cos\theta \\
  \end{array}
  \right].
\end{eqnarray}
The first two equations describe the short sides and
the latter two describe the long ones.

Using equation~\req{eq:irregular.xyparamint}
in equations~\req{eq:irregular.sideparam1}
to \req{eq:irregular.sideparam4},
we respectively have eight possible solutions
(2 for each side):
\begin{eqnarray}
  \label{eq:irregular.tsol1}
  t = \frac
  {-(y_0 + r\cos\theta) \pm a(x_0 - r\sin\theta)}
  {r/a\sin\theta \mp r\cos\theta},
  \\
  \label{eq:irregular.tsol2}
  t = \frac
  {-(y_0 - r\cos\theta) \pm a(x_0 + r\sin\theta)}
  {r/a\sin\theta \mp r\cos\theta},
  \\
  \label{eq:irregular.tsol3}
  t = \frac
  {-1/a(y_0 + r/a\sin\theta) \pm (x_0 + r/a\cos\theta)}
  {r/a\cos\theta \mp r\sin\theta},
  \\
  \label{eq:irregular.tsol4}
  t = \frac
  {-1/a(y_0 - r/a\sin\theta) \pm (x_0 - r/a\cos\theta)}
  {r/a\cos\theta \mp r\sin\theta}.
\end{eqnarray}
Each of these has to be tested to see if $-1 \le t \le 1$.
If so, then upon substituting the values of
$t$ determined in
in equations~\req{eq:irregular.tsol1}
through \req{eq:irregular.tsol4} into
equations~\req{eq:irregular.sideparam1}
through \req{eq:irregular.sideparam4},
we have
\begin{eqnarray}
  \label{eq:irregular.xsol1}
  x = \frac
  {(x_0\sin\theta - y_0\cos\theta - r)}
  {\sin\theta \mp a\cos\theta}, 
  \\
  \label{eq:irregular.xsol2}
  x = \frac
  {(x_0\sin\theta - y_0\cos\theta + r)}
  {\sin\theta \mp a\cos\theta}, 
  \\
  \label{eq:irregular.xsol3}
  x = \frac
  {(x_0\cos\theta + y_0\sin\theta + r/a)}
  {\cos\theta \pm a\sin\theta}, 
  \\
  \label{eq:irregular.xsol4}
  x = \frac
  {(x_0\cos\theta + y_0\sin\theta - r/a)}
  {\cos\theta \pm a\sin\theta}, 
\end{eqnarray}
where $y= \pm ax$ in all cases.

We are now able to write down 
how to determine the size of the largest
rectangle that may fit in given it is centered
at the origin with an angle $\phi$ and
there is one other rectangle already
present at $(x_0,y_0)$ oriented
at an angle $\theta$.

\begin{enumerate}
\item Determine whether or not the new rectangle lies
  within the old one using equation~\req{eq:irregular.recteq3}.
\item If it is not enclosed,
  rotate the coordinate system by an angle
  $\phi$ so the central rectangle sits square with
  the axes.  All that needs to be done is to
  move the existing rectangle from $(x_0,y_0)$
  to
  $(\cos\phi x_0 + \sin\phi y_0,-\sin\phi x_0 + \cos\phi y_0)$,
  and change its angle of rotation to
  $\theta' = \theta-\phi\pmod \pi$.
\item Find the position of the four corner points
  using equation~\req{eq:irregular.fourcorners}
  with $\theta'$ and $(x_0',y_0')$.
\item Determine which, if any, intersection points with the diagonals
  of the new rectangle exist by calculating $t$ in
  equations~\req{eq:irregular.tsol1}
  through \req{eq:irregular.tsol4}.
\item Each value of $t$ that satisfies $-1 \le t \le 1$,
  is a valid intersection point.
  The positions of all such points 
  are then taken from equations~\req{eq:irregular.xsol1}
  through~\req{eq:irregular.xsol4} along with $y=\pm a x$.
\item For each valid $(x,y)$ pair,
  calculate $\max(|ax|,|y|)$ which gives us 
  the half-width of the shortest side of
  the rectangle centered at the origin that passes
  through the point $(x,y)$.  Take
  the minimum over all such values to find
  the half-width of the shortest
  side of the actual largest rectangle that may be inserted
  without overlapping.
\end{enumerate}

\end{document}